\documentclass[letterpaper, 10 pt, conference]{ieeeconf}  
\IEEEoverridecommandlockouts
\usepackage{cite}
\usepackage{amsmath,amssymb,amsfonts}
\usepackage[ruled,vlined]{algorithm2e}
\usepackage{gensymb}
\usepackage{caption}
\usepackage{siunitx}
\usepackage{graphicx}
\usepackage{booktabs}
\usepackage{textcomp}
\usepackage{mathtools}
\usepackage{xcolor}
\usepackage{hyperref}
\usepackage{todonotes}
\usepackage{amsfonts}
\usepackage{amssymb}
\usepackage{amsmath}
\usepackage{hyperref}
\usepackage{xcolor}
\usepackage{user_macros}
\usepackage{soul}
\usepackage{diagbox}
\usepackage[nolist,nohyperlinks]{acronym}
\def\BibTeX{{\rm B\kern-.05em{\sc i\kern-.025em b}\kern-.08em
    T\kern-.1667em\lower.7ex\hbox{E}\kern-.125emX}}
    
\newcommand{\sgnote}[1]%
{\textcolor{green}{\textbf{Note: #1}}}
\newcommand{\shnote}[1]%
{\textcolor{blue}{\textbf{Note: #1}}}

\usepackage{siunitx} 

\begin{document}
\bstctlcite{IEEEexample:BSTcontrol}
\begin{acronym}
\acro{HJ}{Hamilton-Jacobi}
\acro{HJI}{Hamilton-Jacobi-Isaacs}
\acro{ODE}{Ordinary Differential Equation}
\acro{MPC}{Model Predictive Control}
\acro{MDP}{Markov Decision Processes}
\acro{RMSE}{Root Mean Squared Error}
\acro{MSE}[MSE]{mean-squared-error}
\acro{RL}{Reinforcement Learning}
\acro{PDE}{Partial Differential Equation}
\acro{ASV}{Autonomous Surface Vehicles}
\acro{brs}[BRS]{backward reachable set}
\acro{BRT}{backward reachable tube}

\acro{HJ-MTR}{Hamilton-Jacobi Multi-Time Reachability}
\acro{GPGP}{Great Pacific Garbage Patch}
\acro{ctrl1}[Floating]{Passive Floating Controller}
\acro{ctrl2}[MTR-no-Obs]{MTR with no obstacles controller}
\acro{ctrl3}[Switch-MTR-no-Obs]{switching controller}
\acro{ctrl4}[MTR]{multi-time HJ reachability closed loop controller with obstacles}
\acro{ctrl5}[Switch-MTR]{switching controller with obstacles}
\acro{ctrl6}[SmallDist-MTR]{MTR with small disturbance}
\acro{CBF}{Control Barrier Function}
\acro{COLREGS}[COLREGS]{Convention on the International
Regulations for Preventing Collisions at Sea}
\acro{TTR}{Time-to-Reach}
\end{acronym}

\title{\Large \bf
Stranding Risk for Underactuated Vessels in Complex Ocean Currents: \\ Analysis and Controllers}

\author{Andreas Doering$^{1,2,*}$, Marius Wiggert$^{1,*}$, Hanna Krasowski$^{2}$, Manan Doshi$^{3}$\\ Pierre F.J. Lermusiaux$^{3}$ and Claire J. Tomlin$^{1}$
\thanks{$^{*}$ A.D. and M.W. have contributed equally to this work. }
\thanks{$^{1}$ A.D., M.W., and C.J.T. are with the Department of Electrical Engineering and Computer Sciences, University of California, Berkeley, USA. For inquiries contact: {\tt\small mariuswiggert@berkeley.edu}}
\thanks{$^{2}$ A.D. and H.K. are with the School of Computation, Information and Technology of the Technical University of Munich, Germany}
\thanks{$^{3}$ M.D. and P.F.J.L. are with the Department of Mechanical Engineering at the Massachusetts Institute of Technology, USA.}
\thanks{The authors gratefully acknowledge the support of the C3.ai Digital Transformation Institute, the IFI fellowship
of the German Academic Exchange Service (DAAD) funded by the Federal Ministry of Education and Research (BMBF)
, the research training group ConVeY funded by the German Research Foundation under grant GRK 2428, the DARPA Assured Autonomy Program, and the ONR BRC program.}
}
\maketitle

\begin{abstract}
Low-propulsion vessels can take advantage of powerful ocean currents to navigate towards a destination. Recent results demonstrated that vessels can reach their destination with high probability despite forecast errors.
However, these results do not consider the critical aspect of safety of such vessels: because of their low propulsion which is much smaller than the magnitude of currents, they might end up in currents that inevitably push them into unsafe areas such as shallow areas, garbage patches, and shipping lanes. In this work, we first investigate the risk of stranding for free-floating vessels in the Northeast Pacific. We find that at least 5.04\% would strand within 90 days. Next, we encode the unsafe sets as hard constraints into Hamilton-Jacobi Multi-Time Reachability (HJ-MTR) to synthesize a feedback policy that is equivalent to re-planning at each time step at low computational cost. While applying this policy closed-loop guarantees safe operation when the currents are known, in realistic situations only imperfect forecasts are available. 
We demonstrate the safety of our approach in such realistic situations empirically with large-scale simulations of a vessel navigating in high-risk regions in the Northeast Pacific. We find that applying our policy closed-loop with daily re-planning on new forecasts can ensure safety with high probability even under forecast errors that exceed the maximal propulsion. Our method significantly improves safety over the baselines and still achieves a timely arrival of the vessel at the destination. 
\end{abstract}
\section{Introduction}
\label{sec:introduction}
Autonomous systems are increasingly deployed for long-term tasks and need to operate energy-efficient. For systems operating in the oceans or in the air, this leads to a growing interest in utilizing the dynamics of the surrounding flows as a means of propulsion. Stratospheric balloons and airships utilize wind fields\cite{bellemare2020autonomous, manikandan2021research}, while ocean gliders and active drifters exploit ocean currents \cite{lermusiaux_et_al_springer_HOE2016,meyer2016glider, molchanov2015active,lermusiaux_et_al_TheSea2017}. 

\begin{figure}[h!]
    \centering
    \includegraphics[width=.48\textwidth]{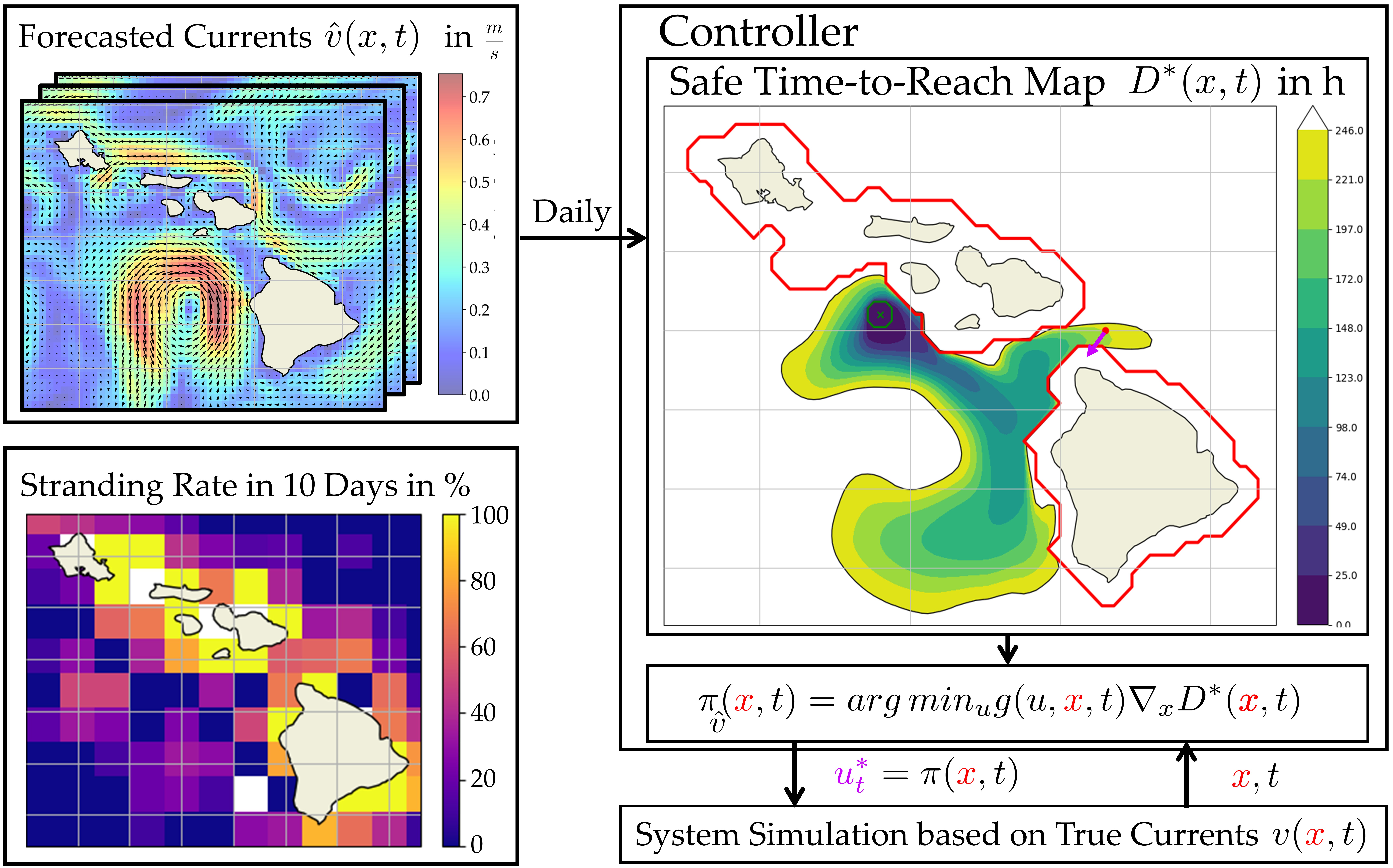}
    \caption{\small
    Our method for safe navigation is based on re-planning on two timescales: 1) compute safe Time-to-Reach map $D^*$ daily as new forecasts become available; 2) For every timestep, e.g., 10 min, we extract {\color{violet}$\bu^*$} from $\pi({\color{red}\vect{x}},t)$, which is a policy equivalent to re-planning. This is necessary, as the real currents $v({\color{red}\vect{x}}, t)$ differ from the forecasted currents $v_t \neq \hat{v_t}$, thus, we will be in a different spatial state ${\color{red}\vect{x}}$ than predicted. Further motivation is provided in the lower left image which displays the stranding rate of free-floating vessels over 10 days.
    }
    \label{fig:head_fig}
\end{figure}

Our recent work \cite{wiggert_etal_CDC2022} has demonstrated that a vessel with just $\SI{0.1}{\meter \per \second}$ propulsion can navigate reliably to a target region by \textit{hitchhiking} on ocean currents of up to $\SI{2}{\meter \per \second}$. This work has further been extended to the application of floating farms which maximize the growth of seaweed over longtime horizons \cite{killer_CDC2023} and to multiple agents that want to stay in proximity to each other to stay connected in local communication networks \cite{hoischen_CDC2023}.

However, these approaches do not factor in safety aspects, although the use of \ac{ASV} in unmanned and long-term operations may pose crucial safety risks. In the event of significant damage, the \ac{ASV} may become inoperable and may be abandoned or sunk, resulting in financial losses and potential environmental impacts. 
One important safety hazard are shallow waters, especially near strong currents, as the \ac{ASV} can easily strand. Another significant safety hazard is entering a garbage patch that has a high concentration of marine debris, which can cover an area of up to 1.6 million \SI{}{\kilo\meter\squared} \cite{lebreton_evidence_2018} as in the case of the \ac{GPGP}. The garbage can get entangled in the \ac{ASV} rotors or damage other components, resulting in loss of control.
Furthermore, collisions with other vessels may cause damage to the \ac{ASV} and potentially endanger the crew of the other vessel. Shipping lanes are another area of increased risk to the \ac{ASV} as they are used by large, fast-moving vessels.
Next, we present related work into safe motion planning for autonomous vessels in maritime environments which can mitigate many of these problems.

\paragraph{Related Work}
The current research focuses on collision avoidance \cite{Geng.2019, Junmin.2020} and compliance with the \ac{COLREGS} \cite{IMO.1972, Meyer.2020, Zhao.2019a, Zhao.2019b, Johansen.2016, Lazarowska.2016} as safety aspects for motion planning of autonomous vessels.
For example, Zhao et al.~\cite{Zhao.2019a} use reinforcement learning to achieve COLREGS-compliant motion planning for encounters with multiple vessels. The studies \cite{Geng.2019, Junmin.2020} only look at collision avoidance and achieve this by employing velocity obstacles. In general, this research on safe motion planning for autonomous vessels considers vessels that are fully actuated. Since our vessel has restricted maneuverability due to its underactuation, it is unnecessary to comply with rules for power-driven vessels from the \ac{COLREGS}. Thus, we consider the safety specification of collision avoidance with largely static obstacles such as shallow areas, shipping lanes, or garbage patches. 

Agents operating in three-dimensional flows can evade obstacles or strong currents by utilizing the third dimension 
which has been demonstrated for stratospheric balloons by \cite{loon.2021}. They ensure safe paths by formulating the problem as a discretized Markov Decision Process and a heuristic cost function. This only ensures heuristic safety and relies on a realistic uncertainty distribution of stratospheric winds, which are not available for ocean currents.

While many papers on maritime safety consider underactuated vessels, most consider underactuation due to non-holonomic actuation of vessels such as \cite{moe_set-based_2017, wu_augmented_2022, liu_collision_2021, borhaug_integral_2009}. In this paper, we define underactuated as having maximum propulsion that is less than the magnitude of error of the forecasted flows, posing severe challenges for the safety of \ac{ASV}s.
The avoidance of dynamic obstacles and forbidden regions including the coordinations of vehicles has been treated using Hamilton-Jacobi Reachability \cite{lolla_et_al_OM2015} and applied in real-time with underwater vehicles to avoid too shallow areas \cite{subramani_et_al_Oceans2017}. 

Robust \ac{MPC} approaches can guarantee safety under disturbances by ensuring that the system is always in a state from which it can reach a robust control invariant set within a finite time horizon \cite{Bemporad.1999, Lazar.2008}. In this robust control invariant set, there always exists a control input that ensures that the system can stay in this set indefinitely. However, in our problem setting with underactuated vessels and imperfect, deterministic ocean current forecasts, no such control invariant set exists, hence robust control with realistic bounds is infeasible.

Our paper makes two main contributions: 
First, we perform an empirical evaluation of stranding risk for free-floating vessels in the Northeast Pacific. 
Second, we present our methods of \ac{HJ-MTR} with re-planning on two timescales for safe motion planning of underactuated \ac{ASV} in a setting with realistic ocean currents and daily forecasts. 
Furthermore, we evaluate our controller with several baseline controllers over a large set of simulated missions.

We structure the paper as follows: we define our problem statement in Sec.~\ref{sec:ProblemStatement} and motivate the need for a safety controller with a stranding study in Sec.~\ref{sec:StrandingStudy}. In Sec.~\ref{sec:method} we introduce our method and summarize \ac{HJ-MTR}. We present experiments we conducted in Sec.~\ref{sec:experiments} and discuss them in Sec.~\ref{sec:discussion}. We conclude and present future work in \ref{sec:conclusion}.
\section{Problem Statement}
\label{sec:ProblemStatement}
We now define the problem of collision avoidance for underactuated vessels by introducing the flow model, vessel model, and representation of obstacles.  We then introduce the notion of stranding as our key performance measure.
\subsection{System Dynamics, Obstacles and Target}
We consider moving in a general time-varying non-linear flow field $v(\x,\t) \to \mathbb{R}^n$, with $\x \in \mathbf{R}^n$ representing the spatial state, 
$\t \in \left[0,T\right]$ the time and $n$ the dimension of the spatial domain. In our case, $n=2$ as we regard an \ac{ASV} operating on the ocean surface.
We denote the actuation signal by $\bu$ from a bounded set  $\mathbb{U} \in R^{n_u}$ with $n_u$ the dimension of the control. 
Let $\stt(\s) \in \mathbf{R}^n$ denote the position of our ASV at time $\s$. Our model for the dynamics of the ASV is given by 
\begin{align}
\label{eq:continuous_general_dynamical_system}
\small
    \dot{\stt}(\s) &= \vect{f}(\stt(\s), \bu(\s), \s)\\
    &= v(\stt,\s) +  g(\bu, \vect{x}, t) \quad \forall  \s \in [0, \T],
\end{align}
Here, the actuation $\bu$ corresponds to the relative velocity $g(\bu, \vect{x}, t)$ of the ASV with respect to the ocean. Hence the absolute velocity of the vehicle is given by the vector sum of the ocean currents at the location of the vehicle and the relative velocity of the vehicle with respect to the currents. The maximum actuation of the vessel is constrained by $||g(\bu, \vect{x}, t)||_2 \leq \bu_\text{max}$.
We define the target and obstacle as sets $\tgt  \in \mathbb{R}^n$ and $\obs  \in \mathbb{R}^n$ respectively.
We assume that these sets are not time-dependent. However, note that the methods presented can be extended to time-dependent cases by the algorithm described in \cite{doshi_etal_CDC2022}.
%
\subsection{Problem Setting}
The agent's goal is to navigate \textit{safely} and \textit{reliably} from a start state $\vect{x}_0$ at start time $t_0$ to a target region $\tgt  \in \mathbb{R}^n$. 
We employ the same empirical definition of \textit{reliability} as \cite{wiggert_etal_CDC2022} defining it as the success rate of a controller navigating from $\vect{x}_0$ at $t_0$ to $\tgt$  within a maximum allowed time $T_{max}$, over a representative set of start-target missions $\{\vect{x}_0, t_0, \tgt\} \in \mathbb{M}$.
We define stranding as an agent entering the obstacle set $\obs$ before $T_{max}$. We then quantify \textit{safety} as the stranding rate of a controller over the same set of missions. 
We define stranding as entering waters with depth less than \SI[round-pad = false]{-150}{\meter}. This is application specific and includes entering obstacles such as garbage patches or areas with high traffic density. 

We use the oceanographic systems of HYCOM \cite{chassignet2009us} and Copernicus \cite{CopernicusGlobal}, similar to prior reasearch \cite{wiggert_etal_CDC2022}. The systems each offer a 5-10 day ocean current forecast based on their models with daily updates. They also offer a so-called hindcast with higher accuracy, which is assimilated from further data and published several days later. 
For realisitc simulation of real operations, the forecasted currents $\hat{v}$ received by the vessel need to differ from the true currents $v$ that are used by the simulation by the forecast error $\delta$, which needs to be comparable to empirical forecast errors of the oceanographic system.
If the true currents are known {\textit a priori} and there exists a trajectory that prevents stranding, our method guarantees safety. However, we are interested in realistic settings with a complex empirical distribution of forecast errors $\delta(\vect{x}, t)$ and severe underactuation e.g. in our experiments
$||g(\bu, \vect{x}, t)||_2=\SI{0.1}{\meter \per \second} \ll \text{RMSE}(\delta) \approx \SI{0.2}{\meter \per \second}$ and currents $||max(v)||_2 \approx  \SI{1.4}{\meter \per \second}$ where safety despite despite worst case  forecast errors is impossible.
Hence, in Sec. \ref{sec:experiments} we evaluate the performance of our method empirically over a large set of missions $\mathbb{M}$ in realistic settings with currents and forecasted currents similar to realistic currents and forecasts $\mathbb{V}$. We evaluate the performance based on the identity functions $\mathbb{I}_{Suc}, \mathbb{I}_{Obs}$ that evaluate to $1$ if in the $\tgt$ and $\obs$ sets respectively and to $0$ otherwise:

\begin{gather}
\small
    \mathbb{E}_{
    \underbrace{\ist, \t \sim \mathbb{M}}_{\substack{\text{initial condition}}};\;
    \underbrace{v, \hat{v} \sim \mathbb{V}}_{\substack{\text{real and forecasted} \\ \text{ocean currents}}}}
    \underbrace{\left\{\mathbb{I}_{Suc}, \mathbb{I}_{Obs}\right\}}_{\substack{\text{Success and Stranding}}}
\end{gather}
\section{Stranding Study}
\label{sec:StrandingStudy}
To illustrate the need for our safety controller, we analyze the rate of stranding for free-floating vessels off the coast of California and Mexico between N\ang{15} and N\ang{40} and W\ang{160} and W\ang{105}.
We define entering an area with a depth of less than \SI{150}{\meter} as stranding, due to operational needs of some \ac{ASV}.

\begin{figure}
    \vspace{4pt}
    \includegraphics[width=.48\textwidth]{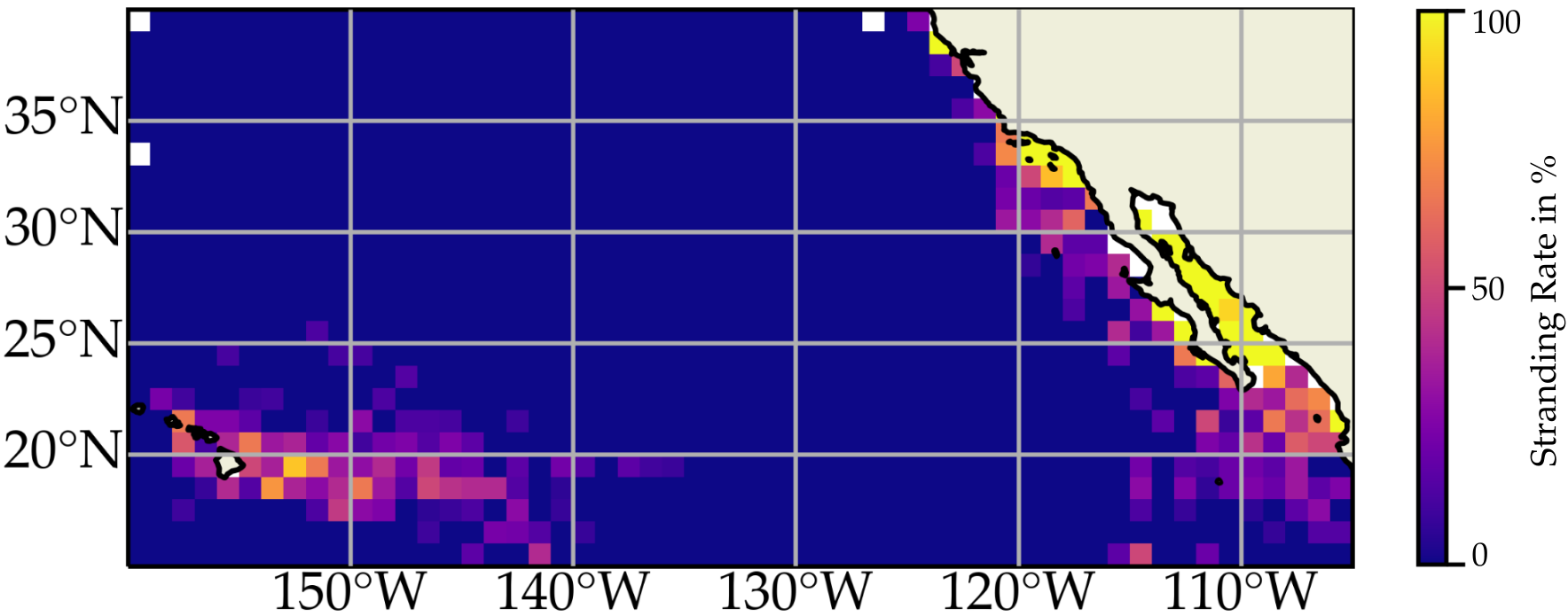}
    \caption{\small
    Rate of free-floating vessels stranding, which we define as entering waters with depth less than \SI[round-pad = false]{-150}{\meter}. 5.04\% of \num[round-pad = false]{10000} simulated vessels starting in the region strand within 90 days. Some vessels float over \ang{20} before stranding.
    }
    \label{fig:stranding}
\end{figure}

The stranding study can be conducted either analytically or experimentally. 
Here, we perform it empirically by sampling \num[round-pad = false]{10000} missions. Each mission consists of a uniformly sampled starting location in the region investigated, outside of the stranding area, and a uniformly sampled starting time. We simulate the trajectories over a time horizon of 10 and 90 days, using Copernicus data for the year 2022. We observe that 1.67\% missions strand within 10 days and 5.04\% strand within 90 days. In Figure \ref{fig:stranding} we show the spatial distribution of the stranding rate over 90 days. We can only count strandings that occur in the defined region, as we stop the simulation for platforms leaving the regarded region, which occurred for 1.78\% missions within 10 days and 18.75\% within 90 days.
\section{Safe HJ Controller}
\label{sec:method}
\label{subsec:method_background}
The forecasts provided by the available ocean forecasting systems are deterministic. This prohibits us from applying probabilistic methods that would require a realistic distribution of currents \cite{lermusiaux_et_al_O2006b}.
Ocean current forecasts exhibit a distribution shift to real currents, e.g., for HYCOM data \cite{chassignet2009us} the global forecast error for speed is $\ac{RMSE}(\delta) = \SI{0.2}{\meter \per \second}$ with a vector correlation \cite{crosby1993proposed} decreasing over the 5-day forecast horizon \cite{metzger2020validation}. For an underactuated vessel with $||\text{max}(g(\bu,\vect{x},t))||_2=\SI{0.1}{\meter \per \second}$, safe navigation cannot be robust against a disturbance of $d=\SI{0.2}{\meter \per \second}$.
Hence we choose to not use robust control 
but to ensure safety despite forecast errors by re-planning on two timescales. First, we compute the value function daily for every forecast we receive using \ac{HJ-MTR} \cite{doshi_etal_CDC2022}. Second, for every time step, e.g. \SI{10}{\minute}, we re-plan by taking the spatial gradient of the value function at $\vect{x}$ to obtain a time-optimal control. This is necessary because the real currents differ from the forecasted currents $v_t \neq \hat{v_t}$, thus, we will be in a different spatial state $\vect{x}$ than predicted. 
\subsection{Multi-Time HJ Reachability for Closed-Loop Control}\label{subsec:closed_loop}
We develop a controller from the theoretic approach of \ac{HJ-MTR}, which has been derived in \cite{doshi_etal_CDC2022}. For completeness, we summarize the technique here.

We first define a modified dynamical system $f_a$ such that the state $\x$ of the vessel becomes frozen when it hits either the target or an obstacle.
\begin{align}
\label{eq:aug_sys_dyn}
\small
    \dot{\stt} = f_a(\stt,\bu,\s) = \begin{cases}
    0, &\quad \stt \in \obs \cup \tgt\\
    v(\stt,\s) +  \bu(\s), &\quad \text{otherwise}.
    \end{cases}.
\end{align}
We define the modified loss function $\rng(\stt,\s)$ such that we earn a reward based on how early we reach the target:
\begin{align}\label{eq:running_cost}
\small
 \rng(\stt,\s) = \begin{cases}
 -\alpha,  &\stt \in \tgt \text{ and } \stt \not\in \obs_\s\\
 0, &\text{otherwise}.
 \end{cases}
\end{align}
Finally, we define the terminal cost function $\trm(\stt)$ to be infinitely high if the \ac{ASV} terminates in an obstacle and is equal to the distance from the target set otherwise.
\begin{align}\label{eq:terminal_cost}
\small
 \trm(\stt) = \begin{cases}
 \infty, & \stt \in \obs_\T\\
 d(\stt,\tgt), & \text{otherwise}.
 \end{cases}.
\end{align}
Finally, we obtain the Hamilton-Jacobi \ac{PDE} which lets us solve for the value function: 
\begin{align}
\label{eq:multitime_hjb}
\small
   \pfracp{\valfunc}{\t} &=
   \begin{cases}
   \alpha,
   & \ist(\t) \in \tgt \cap (\obs_\t)^c\\
   0,
   & \ist(\t) \in \obs_\t\\
   -v(\x,t)- \bu_\text{max}\norm{\Dx{\svalfunc}},
   & \text{otherwise}.
   \end{cases} \nonumber\\
    \svalfunc(\ist, T) &= \begin{cases}
 \infty, & \ist \in \obs_\T\\
 d(\ist,\tgt), & \text{otherwise}.
 \end{cases}. 
\end{align}

\noindent This value function $J^*$ subsequently allows us to compute a feedback policy for this system given by
\begin{align*}
\label{eq:policy}
\small
    \pi_{\hat{v}}(\ist,\t) &= \argmin_{\bu}\left[\Dx{\valfunc} 
    \cdot f(\ist,\bu,\t)\right]\; \forall \ist \not\in \left(\obs_\t \cup    \right)\,,\\
    &= -\frac{\Dx{\valfunc}}{\norm{\Dx{\valfunc}}_2}\bu_\text{max}
\end{align*}

\begin{figure}[h]
    \vspace{4pt}

\includegraphics[width=.42\textwidth]{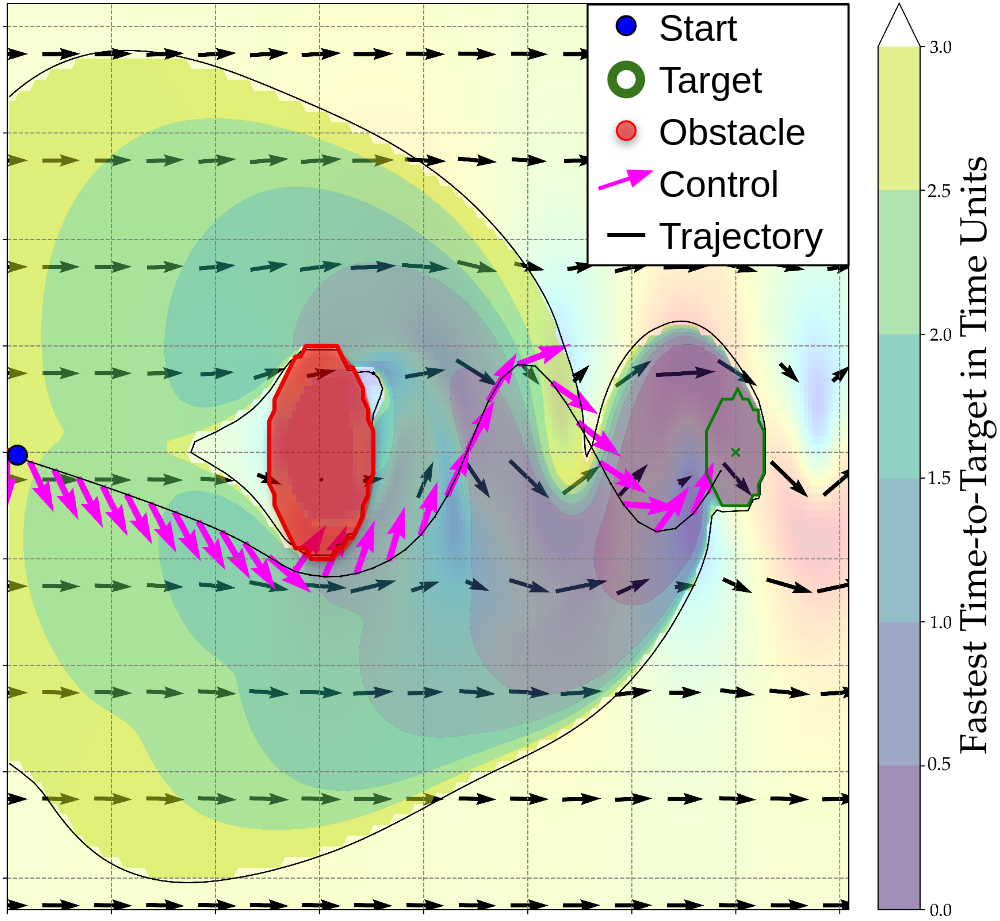}
    \caption{\small 
    Trajectory of our method in analytical currents evading obstacle. The safe Time-to-Reach map spares out obstacle and areas where an underactuated agent is inevitably pushed into the obstacle.}
    \label{fig:analytical}
\end{figure}
This policy guarantees safety when the value function was computed with the true currents $v$. However, in realistic settings with only forecasts $\hat{v}$ available, we apply $\pi_{\hat{v}}$ closed-loop which is equivalent to replanning at every time step.  Applying this policy in closed-loop (see Fig. \ref{fig:head_fig}) means we take the time-optimal control at each state, which is equivalent to full time horizon replanning at each time step. 
We introduce the safe \ac{TTR} map $D^*$, which is easier to interpret and can be easily computed from the value function.  
\begin{align}
    D^*(\vect{x}, t) = T + J^*(\vect{x}, t) -t, \, \forall(\vect{x}, t)\text{ s.t., } J^*(\vect{x},t) \leq 0
\end{align}
We illustrate the interpretability of the safe \ac{TTR} in Fig. \ref{fig:analytical}. If $D^*(\vect{x},t)$ is e.g. $3$, it means that a vessel starting at $\vect{x}$ at time $t$ can reach the target in $3$ time units when following the optimal control (Eq. \ref{eq:policy}).
We solve the \ac{HJ-MTR} in periodic intervals to update the safe \ac{TTR} value function $D^*$. In our work, we solve it once per day upon receiving new forecasts similar to \cite{wiggert_etal_CDC2022}.
This needs to be done, when the next state is different than the predicted state, which is likely due to the forecast error $\delta$.

In summary, there are three core advantages of our method compared to classical \ac{MPC} with non-linear programming. We can guarantee time-optimality in non-linear dynamics over the full time horizon. We require very low online computation to extract the gradient at each step. In case the vessel cannot reach the destination, the optimal control will attempt to minimize the terminal distance to the target, while non-linear programming would not provide us with a trajectory in such a case.

\section{Experiments}
\label{sec:experiments}
We conduct experiments to ascertain that our method of using \ac{HJ-MTR} with obstacles is able to reach the target without colliding with obstacles despite forecast errors.
We simulate a large number of missions on realistic ocean currents and compare the performance of our control schema to baseline methods.

\subsection{Experimental Set-Up}
Our experiments investigate the stranding rate and reliability of several controllers for navigating a two-dimensional \ac{ASV} with fixed magnitude, holonomic actuation of $||g(\bu, \vect{x}, t)||_2 = ||\bu||_2 = \SI{0.1}{\meter \per \second}$. The control input is the angle $\theta$ for steering the \ac{ASV} in ocean currents $v(\vect{x},t) \in [\SI{0}{\meter \per \second}, \SI{1.4}{\meter \per \second}]$, which the vessel utilizes to reach its target region. Additionally, we describe how we ascertain the realism of our ocean forecast simulation and the creation of our obstacle sets, and the generation of a representative set of missions. Subsequently, we explain our baseline methods and evaluation metrics.

\paragraph{Realistic Ocean Forecast Simulation}
In a real-world setting, a vessel can receive the most recent forecast in regular intervals, e.g. daily, and provide it to the control methods to perform replanning.
We employ ocean current hindcast data from Copernicus \cite{CopernicusGlobal} and HYCOM \cite{chassignet2009us} for the region off the coast of California and Mexico between N\ang{15} and N\ang{40} and W\ang{160} and W\ang{105}. 

We simulate the system dynamics based on hindcast as the true flow $v(\vect{x},t)$ using Copernicus hindcasts and use a series of 5 days of HYCOM hindcasts as forecasts for planning. It should be noted that, unlike HYCOM, Copernicus incorporates tidal currents into its forecasts.
We want to ensure a realistic simulation of the forecast error $\delta$. This forecast error can be measured with various metrics on the currents such as \ac{RMSE}, vector correlation, and separation distance \cite{CopernicusGlobal, chassignet2009us}. In our simulations,  these are on average \SI{0.18}{\meter \per \second} \ac{RMSE}, which is close to the validation \ac{RMSE} of \SI{0.19}{\meter \per \second} of the HYCOM forecast error \cite{metzger2020validation}. We measured 0.63 vector correlation compared with 0.64 for HYCOM \cite{metzger2020validation} and 0.62 for Copernicus \cite{CopernicusGlobal}, each measured at $t=\SI{71}{\hour}$, with a value of 2 representing perfect correlation and 0 no correlation. Thus, our simulation set-up represents realistic situations well.

\paragraph{Obstacles derived from Bathymetry}
The bathymetry data we employ is the GEBCO 2022 grid \cite{gebco_bathymetric_compilation_group_2022_gebco_2022_2022}. It is a global, continuous terrain grid with a resolution of \ang{;;15}. We coarsen it to the same resolution of the current data \ang{;5} by using the maximum in each grid cell to overestimate the elevation in each grid cell. We further precompute a distance map corresponding to the minimal distance to obstacle areas of depth under \SI{150}{\meter} for each grid cell. This is done by employing a Breadth-First-Search starting in the obstacle area with a distance of zero and exploring outwards. We use this distance as switching condition for the \ac{ctrl3}.

\paragraph{Large Representative Set of Missions}
We generate a representative set of 1146 start-target missions $\mathbb{M}$ with the following procedure. We uniformly sample target points $x_{\tgt, center}$ spatially in the region introduced in \ref{sec:StrandingStudy}.
We reject points with a minimum distance below \ang{0.5} to the boundary of the region, with distance to obstacles below \ang{0.025}. We want to generate missions with a higher risk of stranding, hence we limited the maximum distance to an obstacle to \ang{3}. 
We then uniformly sample final times $t_T$ for a time horizon of 10 days while using data from 2022.
To validate each mission is feasible for \ac{ctrl2}, introduced in \cite{wiggert_etal_CDC2022}, given the true currents $v(\mathbf{x},t)$ we calculate the \ac{BRT} using \ac{HJ-MTR} from $x_{\tgt_\text{center}}$  at $t_\tgt$. We sample a start position from the \ac{BRT} so that the \ac{ASV} can reach the target within 5-9 days. Finally we define the target region $\tgt$ to be a circular region with radius $r_\tgt = \text{\ang{0.1}}$ around $x_{\tgt, center}$. Note that the introduction of obstacles increases the difficulty of the mission and renders some of these missions infeasible, as they can block a desired path.

\paragraph{Controllers}
It has been shown that navigating from start to target with an underactuated \ac{ASV} can be successfully done by hitchhiking ocean currents \cite{subramani_et_al_Oceans2017,
wiggert_etal_CDC2022}. Yet their controller \ac{ctrl2} does not incorporate safety aspects into its value function. Our work seeks to extend its capabilities by incorporating collision avoidance into planning \cite{lolla_et_al_OM2015}. We hence compare the performance of several controllers to the baseline presented in \cite{wiggert_etal_CDC2022}.
The first controller is the \ac{ctrl1}, with an actuation of $\bu = 0$. It is the same we used in \ref{sec:StrandingStudy} and serves the purpose to present a lower bound on the performance and should show how complex it is to safely complete the missions.
The second controller is a reactive safety controller, meaning it does not reason about currents for safety. Instead it is a switching controller with switching condition being the distance to the closest obstacle. If the distance is below a threshold, the safety controller takes over and actuates with full actuation into the direction of the largest distance to obstacles. We set the threshold to \SI{20}{\kilo \meter}. Once the distance to obstacles is larger than the threshold, the navigation controller \ac{ctrl2} takes over again. We utilize it to be able to compare to a simple solution that adds safety functionality to the baseline controller \ac{ctrl2}.
Our \ac{ctrl4} is the controller we presented in detail in \ref{sec:method}. It is the primary contribution of our paper. There are two ablations of the \ac{ctrl4}. The \ac{ctrl5} is a switching controller that uses \ac{ctrl4} for navigation. The switching condition is the same as for the \ac{ctrl2}.
The last controller we examine is the \ac{ctrl6}, which is the \ac{ctrl4} controller with an unrealistically low disturbance of $d=\SI{0.05}{\meter \per \second}$. As explained in \ref{sec:method} in a realistic setting with $d=\SI{0.2}{\meter \per \second}$ we cannot be robust with an actuation of only $\bu_{max}=\SI{0.1}{\meter \per \second}$.

\paragraph{Evaluation Metrics}
We define our key metric \textit{stranding rate} as the rate of a controller entering the obstacle set $\obs$ over the set of missions $\mathbb{M}$. We further evaluate the \textit{reliability}, defined as the success rate of a controller over the set of missions $\mathbb{M}$ \cite{wiggert_etal_CDC2022}. 

\subsection{Experimental Results}
We evaluate the controllers performances over $||\mathbb{M}|| = 1146$ start-target missions and run the simulation for $T_{max}=\SI{240}{\hour}$. If the \ac{ASV} collides with an obstacle, we terminate the mission and count it as stranded, if it reaches the target region within $T_{max}$ we count it as success, if it does neither we count it as timeout.
In complex flows with forecast errors and in close proximity to obstacles, our controller \ac{ctrl4} has a stranding rate of only $0.96\%$, compared to $4.71\%$ of the baseline \ac{ctrl2} (Table \ref{tab:results}).

We evaluate if the stranding rate of our controllers is lower than the baseline of \ac{ctrl2} in a statistically significant manner by performing a one-sided two-sample z proportion test for the other controllers. 
Let $\Gamma$ be the stranding rate of a controller and our null hypothesis be:
\begin{align}
 H_0: \Gamma_{\ac{ctrl2}} = \Gamma_\text{controller}.
\end{align}
With the alternate hypothesis:
\begin{align}
H_A: \Gamma_{\ac{ctrl2}} > \Gamma_\text{controller}.
\end{align}
The stranding rate of both controllers is higher than \ac{ctrl2} in a statistically significant way with p-values \ac{ctrl3} $p=2.6e^{-3}$, \ac{ctrl4} $p=3.1e^{-8}$, \ac{ctrl5} $p=9.3e^{-7}$, \ac{ctrl6} $p=9.5e^{-5}$). Additionally, the success rate of \ac{ctrl4} is not reduced by safety and even shows the highest success rate.

\begin{table}[!htb]\centering
\caption{\small
We compare the performance of multiple controllers, the arrows indicate if high, or low is preferred. Our \ac{ctrl4} significantly outperforms the other controllers with respect to the stranding rate. The $^*$ marks statistically significant lower stranding rates compared to the \ac{ctrl2}.}
    \begin{tabular}{lll} \toprule
     Controller & Stranding Rate $\downarrow$ & Success Rate $\uparrow$ \\ \midrule
    \acs{ctrl4} & \textbf{0.96}\%$^*$ & \textbf{37.26}\% \\
    \acs{ctrl5} & 1.31\%$^*$ & 37.17\%  \\
    \acs{ctrl6} & 1.92\%$^*$ & 33.16\% \\
    \acs{ctrl3} & 2.53\%$^*$ & 36.82\% \\
    \acs{ctrl2} & 4.71\% & 36.91\%\\
    \acs{ctrl1} & 4.89\% &  2.53\%\\
    \bottomrule
    \end{tabular}
    \label{tab:results}
\end{table}
\section{Discussion}
\label{sec:discussion}
We note that our controllers exhibit a lower success rate than in \cite{wiggert_etal_CDC2022}. We believe this is due to three differences in the set-up. First, the time-to-target for each mission in \cite{wiggert_etal_CDC2022} is between 20-120h with $T_{max}=\SI{150}{\hour}$, while our sampled time-to-target is 120-216h with $T_{max}=\SI{240}{\hour}$. 
Hence our missions are longer and have smaller time buffers to reach the target. In extreme cases, their missions are expected to finish in \SI{20}{\hour} with \num{130} additional hours to reach the target before $T_{max}$, while in the worst case, our missions can have a \SI{216}{\hour} mission with a buffer of \SI{24}{h}.
Second, the sampling is feasible for \ac{ctrl2} without considering obstacles (Sec.  \ref{sec:experiments}). Hence missions may be unfeasible for \ac{ctrl2} due to stranding on obstacles and unfeasible for the other controllers, as it may take them time to circumvent obstacles in the path. 
Third, we sample missions with a maximum distance to shore of \ang{3}, exposing the vessels more to tidal currents near shore.

\section{Conclusion and Future Work}
\label{sec:conclusion}
In this work, we have demonstrated that \ac{HJ-MTR} with obstacles can be used to reduce the rate of stranding even in complex flows using daily forecasts with large errors. We evaluated our method over a large set of 5-9 day start-to-target missions distributed spatially near the Coast of California, Hawaii, and the Baja California area and temporally across the year 2022 using realistic ocean currents. In our experiments, our method has achieved a stranding rate of $0.96\%$ which is significantly lower than the baseline controllers and also has a slightly higher success rate. 
While we have demonstrated the ability of our method with two-dimensional ocean currents, we emphasize that it is also applicable in a three-dimensional setting such as underwater or in the air. Furthermore, \ac{HJ-MTR} is able to handle dynamic constraints \cite{doshi_etal_CDC2022}. However, including dynamic obstacles such as ships that move fast and change their course would require a higher frequency of re-planning to account for those changes, resulting in higher computational costs.

In the future, we plan to model zones of a potential hazard, e.g.\ shipping lanes and garbage patches, as soft constraints, where instead of preventing entering altogether it would be beneficial to e.g. minimize the time spent therein.
By reducing time in shipping lanes an \ac{ASV} could avoid many vessels. As of now, it is also uncertain how underactuated \ac{ASV}s would be classified under the \ac{COLREGS} and if evasion is necessary or if they should stop their propulsion to be floated along a vessel \cite{sun_et_al_ACC2017,sun_et_al_JGCD2017}.
Getting the rotors of an \ac{ASV} entangled in the garbage can render it inoperable, hence it is beneficial to avoid areas with a larger density of garbage such as the center of the \ac{GPGP}, while not making the whole 1.6 million \SI{}{\kilo\meter\squared} \cite{lebreton_evidence_2018} of the \ac{GPGP} an obstacle to be avoided. We can investigate using a risk-based extension of a soft-edge and dynamic forbidden region \cite{lolla_et_al_OM2015,subramani_lermusiaux_CMAME2019}.




\bibliographystyle{IEEEtran}
\bibliography{references,mseas,krasowski}

\end{document}